\documentclass[aps,prb,twocolumn,superscriptaddress]{revtex4}
\usepackage[utf8]{inputenc}
\usepackage{amsmath}
\usepackage{amssymb}
\usepackage{mathtools}
\usepackage{graphicx}

\begin{document}

\title{Josephson current between topological and conventional superconductors}

\author{P.\ A.\ Ioselevich}
\affiliation{Max Planck Institute for Solid State Research, Heisenbergstr.\ 1, Stuttgart, 70569 Germany}
\affiliation{L.\ D.\ Landau Institute for Theoretical Physics, Kosygin str.\ 2, Moscow, 119334 Russia}

\author{P.\ M.\ Ostrovsky}
\affiliation{Max Planck Institute for Solid State Research, Heisenbergstr.\ 1, Stuttgart, 70569 Germany}
\affiliation{L.\ D.\ Landau Institute for Theoretical Physics, Kosygin str.\ 2, Moscow, 119334 Russia}

\author{M.\ V.\ Feigel'man} 
\affiliation{L.\ D.\ Landau Institute for Theoretical Physics, Kosygin str.\ 2, Moscow, 119334 Russia}
\affiliation{Moscow Institute of Physics and Technology, Institutsky per.\ 9, Dolgoprudny, 141700 Russia}

\begin{abstract}
We study the stationary Josephson current in a junction between a topological and an ordinary (topologically trivial) superconductor. Such an S-TS junction 
hosts a Majorana zero mode that significantly influences the current-phase relation. The presence of the Majorana state is intimately related with the breaking 
of the time-reversal symmetry in the system. We derive a general expression for the supercurrent for a class of short topological junctions in terms of the 
normal state scattering matrix. The result is strongly asymmetric with respect to the superconducting gaps in the ordinary ($\Delta_0$) and topological 
($\Delta_{\mathrm{top}}$) leads. We apply the general result to a simple model of a nanowire setup with strong spin-orbit coupling in an external magnetic field 
and proximity-induced superconductivity. The system shows parametrically strong suppression of the critical current $I_c \propto \Delta_{\mathrm{top}}/R_N^2$ in 
the tunneling limit ($R_N$ is the normal state resistance). This is in strong contrast with the Ambegaokar-Baratoff relation applicable to junctions with 
preserved time-reversal symmetry. We also consider the case of a generic junction with a random scattering matrix and obtain a more conventional scaling law 
$I_c \propto \Delta_{\mathrm{top}}/R_N$.
\end{abstract}

\maketitle
\section{Introduction}
Topological superconductors and insulators share the property of hosting a gapless edge or surface mode while having a gap in the bulk
spectrum.\cite{reviewQiZhang, reviewHasanKane} The edge mode of a one-dimensional topological superconductor is a Majorana zero mode -- a self-conjugate
excitation with zero energy.\cite{Kitaev01} Majorana modes are half-fermions obeying non-abelian braiding statistics. Their special properties make them
promising ingredients to topological quantum computing.\cite{reviewTopComp}

A lot of effort is put at the moment into the observation of Majorana excitations in condensed matter systems.\cite{reviewAlicea, reviewBeenakker} A major 
route to detecting these modes is by transport measurements. When probed by a tunneling contact, the Majorana mode induces resonant Andreev reflection, leading
to a robust zero-bias peak in the tunneling conductance.\cite{FuKaneZBP, AkhmerovZBP, LawLeeNg, IOF2012, IF2013} Such a peak has been observed in several
experiments \cite{Mourik2012, Das2012, Deng2012} on nanowires with strong spin-orbit interaction and induced superconductivity.\cite{Sarma, OregOppen} The peak
is robust to perturbations and only appears in the topological phase of the system. This allows to identify it with the presence of a Majorana zero mode. 

A different kind of transport experiment involves Josephson junctions between two topological superconductors. Each topological superconductor gives rise to one 
Majorana mode. The pair of Majorana modes from the two leads hybridise into a single Andreev bound state with energy $E\sim |\cos\varphi/2|$, where $\varphi$ 
is the superconducting phase difference at the junction. This bound state is non-degenerate and leads to a parity switch -- the fermionic parity of the systems 
ground state changes as $\varphi$ passes $\pi$. This leads to the celebrated fractional Josephson effect \cite{Kitaev01, Yakovenko, FuKane2009, Sarma, 
OregOppen, IF2011, AliceaOppen, RanHosur} where the current $I(\varphi)$ contains a $4\pi$-periodic component directly attributable to the hybridised Majorana 
state. 

\begin{figure}
\centering
\hspace*{-3pt}\includegraphics[width=0.475\textwidth]{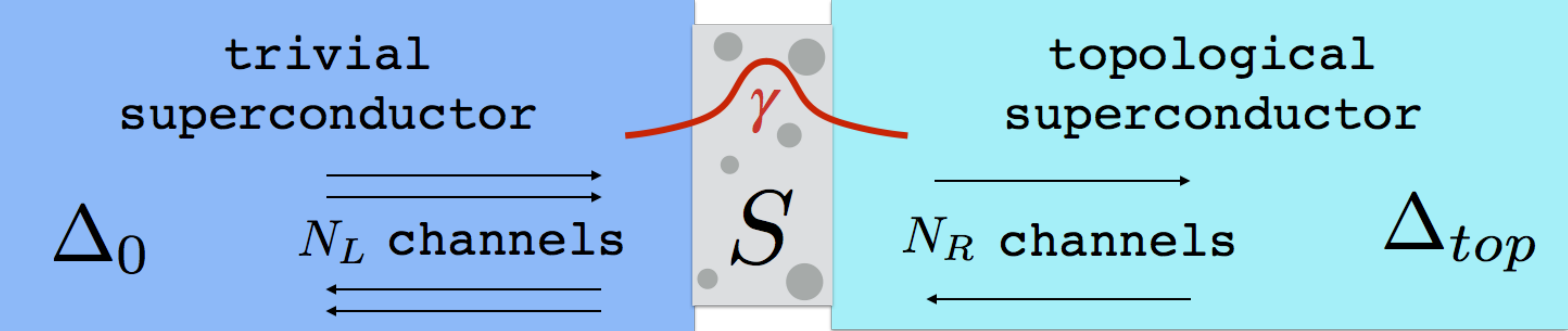}
\caption{(Color online) Schematic of the junction between an ordinary (left) and topological (right) superconductor with $N_L$ and $N_R$ channels, 
respectively. Normal transport in the junction is characterised by the $(N_L + N_R)$-sized scattering matrix $S$. The Josephson junction hosts a 
localised Majorana zero mode $\gamma$. 
}
\label{fig1}
\end{figure}

In this paper we consider a Josephson junction where only one of the leads is a topological superconductor, while the other is topologically trivial, i.e. an 
S-TS junction. Similarly to the N-TS-junction setups used to probe topological superconductivity by normal transport, our setup deals with a single Majorana 
mode. This mode is topologically protected from local perturbations and its energy stays zero at any phase differences $\varphi$. Thus the Majorana zero mode 
does not directly contribute to the supercurrent, unlike what happens in junctions where both superconductors are topological. Nevertheless, as we show below, 
the stationary Josephson current in the S-TS junction is generally quite different from the current in conventional junctions even in the simplest tunnelling
limit.\cite{Ambegaokar}

We describe the junction between two superconductors within the general quasi-one-dimensional formalism in terms of conducting channels. Both superconducting 
leads are represented as wires with broken spin symmetry and induced superconductivity. The distinction between ordinary and topological superconductor is in 
the parity of the number of channels. When a superconducting wire with an odd number of channels is terminated, an odd number of bound states appears at its 
end. One of these bound states is self-conjugate and has exactly zero energy due to the mirror symmetry of the spectrum of the superconducting Hamiltonian. 
This is the Majorana mode characteristic to the topological superconductor. 

We consider a generic junction between two superconductors with the number of channels $N_L$ and $N_R$ in the left and right lead being even and odd, 
respectively. Thus we have an ordinary (topological) superconductor as the left (right) lead of our junction, see Fig.\ \ref{fig1}. Boundary conditions at the 
end of a wire with an odd number of channels necessarily break the time-reversal symmetry $\mathcal{T}$. The same is true for any junction where the parity 
of the number of channels changes. Thus broken time-reversal symmetry is an essential property of the S-TS junction required for an unpaired Majorana mode to 
form.

We derive general expressions for the supercurrent and the spectrum of Andreev bound states for a class of short S-TS junctions in terms of their scattering 
matrix in the normal state. This requires a generalisation of classical results \cite{Beenakker91, Bagwell} for the Josephson current in time-reversal-symmetric 
junctions. While in the latter case all measurable quantities can be expressed in terms of normal transmission probabilities of the device, the former case of 
broken $\mathcal{T}$ symmetry leads to more complicated supertransport that requires taking into account the full scattering matrix. We further apply our 
general results to a particular minimal model with two and one channel in the ordinary and topological superconducting leads, respectively, as shown in Fig.\ 
\ref{fig1}. The expressions for the supercurrent and the excitation spectrum of such an S-TS junction appear to be strongly asymmetric in their dependence on 
the superconducting gaps $\Delta_0$ and $\Delta_\text{top}$ of the ordinary and topological leads.

We further consider the particular case of a tunneling junction described by the one-dimensional Hamiltonian of Refs.\ \onlinecite{Sarma, OregOppen}. It 
corresponds to a nanowire with parabolic spectrum, spin-orbit coupling, and induced superconductivity in an external magnetic field. This model is widely used 
to describe the recent N-TS transport experiments.\cite{Mourik2012, Das2012, Deng2012} We find that the critical current is suppressed in the tunneling limit 
and scales as the square of the normal conductance of the junction. This result is in sharp contrast with the Ambegaokar-Baratoff relation \cite{Ambegaokar} 
that predicts linear scaling for tunnel junctions with $\mathcal{T}$ symmetry. We further explore this discrepancy by considering a generic junction with a 
random scattering matrix. It turns out that the typical Josephson current in such a random junction is parametrically larger and obeys the linear scaling with 
normal conductance similar to the Ambegaokar-Baratoff relation.

The paper is organised as follows. In Section \ref{sec_formalism} we introduce the scattering matrix formalism and derive general relations for the 
supercurrent. In Section \ref{sec_minimal} we solve the minimal model of an S-TS junction with two and one channels in the leads. This result is applied to 
the nanowire with parabolic spectrum in Section \ref{sec_nanowire}. A generic minimal junction with a random scattering matrix is discussed in Section 
\ref{sec_RMT}. We summarise the results in the concluding Section \ref{sec_conclusion}.

\section{$S$-matrix formulation of Josephson current}
\label{sec_formalism}

We consider a Josephson junction between two quasi-one-dimensional conductors with $N_L$ and $N_R$ channels. In the normal state, each channel hosts a pair of 
counter-propagating states related by time-reversal symmetry. We will describe the leads by the quasiclassical Hamiltonian $\hat \xi = -i \sigma_z \hat v\, 
\partial/\partial x$. Here $\hat v$ is the diagonal matrix of Fermi velocities and $\sigma_z$ is a Pauli matrix in the space of left-right propagating modes. 
The antiunitary time-reversal operator has a simple form $\mathcal{T} = i \sigma_y \mathcal{C}$ in the channel basis, where $\mathcal{C}$ denotes complex 
conjugation. The Hamiltonian $\hat \xi$ commutes with $\mathcal{T}$. Let us note that $\mathcal{T}^2 = -1$ corresponding to the symplectic symmetry of the 
wires.

The properties of the junction in the normal state are encoded in the scattering matrix
\begin{align}
 S
  = \begin{pmatrix}
      r_L & t_{LR} \\
      t_{RL} & r_R
    \end{pmatrix}.
 \label{S}
\end{align}
This is a unitary matrix of the size $N_L + N_R$ that relates the amplitudes of outgoing and incoming waves. The two diagonal blocks $r_{L,R}$ of the 
size $N_L$ and $N_R$ contain reflection amplitudes in the left and right lead while the off-diagonal rectangular $N_L \times N_R$ blocks $t_{LR}$ and $t_{RL}$ 
contain transmission amplitudes. Normal transport properties of the junction can be expressed in terms of transmission probabilities $T_i$ -- eigenvalues of 
the matrix $t^\dagger_{LR} t_{LR}$ or $t^\dagger_{RL} t_{RL}$. Unitarity of $S$ ensures that non-vanishing eigenvalues of these two matrices are identical.

Let us note that the time-reversal symmetry is inevitably broken already in the normal state if $N_L$ and $N_R$ have opposite parity. If the junction possesses 
time-reversal symmetry $\mathcal{T}$, its scattering matrix obeys an additional linear constraint $S = -S^T$. However, a skew-symmetric matrix of the odd size 
$N_L + N_R$ necessarily has a zero eigenvalue which is incompatible with the unitarity of $S$.

In the superconducting state, the leads are described by the Bogoliubov-de Gennes Hamiltonian
\begin{equation}
 H
  = \begin{pmatrix}
      \hat \xi & \Delta \\
      \Delta^* & -\hat \xi
    \end{pmatrix}.
 \label{BdG}
\end{equation}
We assume that the order parameter $\Delta$ takes two different values $\Delta_0$ and $\Delta_\text{top}$ in the ordinary (left) and topological (right) leads 
of the junction. The supercurrent can be expressed as the derivative of the junction free energy $F$ with respect to the superconducting phase difference 
$\varphi$ via the fundamental relation $I = (2e/h) \partial F/\partial\varphi$. In terms of the Matsubara Green function, this can be represented as
\begin{align}
 I
  = -\frac{e}{h}\, \frac{\partial}{\partial\varphi} T \sum\limits_\omega i \omega \int dx\, \mathop{\mathrm{Tr}} G_\omega(x,x).
 \label{IF}
\end{align}
Here Matsubara frequencies take the discrete values $\omega = 2\pi T(n + 1/2)$ and $T$ is the temperature. The Matsubara Green function is a matrix in channel 
and Nambu space satisfying the equation $(i\omega - H) G_\omega(x,x') = \delta(x-x')$ with the Hamiltonian (\ref{BdG}). Explicitly solving this 
equation,\cite{BrouwerBeenakker} we relate the Josephson current to the normal state scattering matrix $S$, Eq.\ (\ref{S}),
\begin{align}
 I
  = -\frac{e}{h}\, \frac{\partial}{\partial\varphi} T \sum\limits_\omega \ln \det \bigl[ 1 - S_N S_A(\omega) \bigr].
 \label{itodet}
\end{align}
Here the determinant of the size $2N_L + 2N_R$ contains the matrices of normal and Andreev scattering
\begin{gather}
 S_N
  = \begin{pmatrix}
      S & 0 \\
      0 & -S^*
    \end{pmatrix}, \qquad
 S_A
  = -i\begin{pmatrix}
      0 & r_A \\
      r_A^* & 0
    \end{pmatrix}, \\ \intertext{where}
 r_A
  = \begin{pmatrix}
      e^{-\theta_L+i\varphi_L} & 0 \\
      0 & e^{-\theta_R+i\varphi_R}
    \end{pmatrix}, \quad
 \theta
   = \mathop{\mathrm{arcsinh}} \frac{\omega}{\Delta}.
\end{gather}
The Andreev scattering amplitude $r_A$ takes two different values for the left and right lead allowing for different superconducting gaps and phases. In 
general, the normal scattering matrix $S$ is a function of energy, for simplicity we focus on the limit of a short junction where this energy dependence can be 
neglected.

We will denote the determinant in Eq.\ (\ref{itodet}) by $D$. Nambu structure can be eliminated from this determinant leading to
\begin{align}
 D(\varphi, \omega)
  = \det\left(S - r_A^* S^T r_A\right).
 \label{detD}
\end{align}
Here $\varphi = \varphi_L - \varphi_R$ and an unobservable phase-independent factor is omitted. The phase difference $\varphi$ and Matsubara energy $\omega$ 
enter $D$ through $r_A$ while $S$ contains all the junction specifics. The expression (\ref{itodet}) with the determinant (\ref{detD}) relates the Josephson 
current in an arbitrary short junction to its normal transport characteristics.
\begin{align}
 I
  = -\frac{e}{h}\, \frac{\partial}{\partial\varphi} T \sum\limits_\omega \ln D(\varphi, \omega).
 \label{ID}
\end{align}

Let us briefly review the case of conventional Josephson junctions, where time-reversal symmetry is present in the normal state. In terms of the $S$-matrix, 
$\mathcal{T}$ symmetry means $S = -S^T$. This allows to block-diagonalise the matrix in Eq.\ (\ref{detD}) and express the Josephson current in terms of normal 
transmission probabilities $T_i$. For equal superconducting gaps $\Delta_L = \Delta_R = \Delta$ this leads to the classical short junction result 
\cite{Beenakker91}
\begin{align}
I=-\frac{e}{h}\frac{\partial}{\partial\varphi}\sum\limits_i E_i\tanh\frac{E_i}{2T},\label{BeenakkerI}\\
E_i=\Delta\sqrt{1-T_i\sin^2\varphi/2}.\label{BeenakkerE}
\end{align}

We now return to the S-TS junction. The total number of channels $N_R + N_L$ is odd and the time-reversal symmetry is broken. This prevents us from 
block-diagonalising the matrix in Eq.\ (\ref{detD}) in a general case. The current is determined by the whole scattering matrix rather than by normal transmission 
probabilities only.

Let us demonstrate that the junction hosts a Majorana zero mode. Consider the determinant (\ref{detD}) at zero energy. Using $r_A r_A^*|_{\omega = 0} = 1$, we 
can rewrite it as $D(\varphi, 0) = \det [(r_A S)^T-r_A S]_{\omega = 0}$. The argument of the determinant here is an odd-sized skew-symmetric matrix hence 
$D(\varphi, 0) = 0$. We thus conclude that the junction hosts an excitation with exactly zero energy when $N_L$ and $N_R$ have opposite parity. In particular, 
$N_L = 0$, $N_R > 0$ represents a terminated superconducting wire, hosting a Majorana edge mode whenever $N_R$ is odd. This once again proves that a 
topologically non-trivial (trivial) superconductor is a superconducting nanowire with an odd (even) number of channels.

\section{Minimal model of an S-TS junction}
\label{sec_minimal}

The simplest model of an S-TS junction involves three channels in total with $N_L=2$ and $N_R=1$, see Fig.\ \ref{fig1}. The $S$ matrix (\ref{S}) 
is $3\times3$ in this case with $r_L$ being a $2\times2$ block and $r_R$ being just a complex number. Such a unitary matrix contains in total nine parameters. 
However, all observable quantities are insensitive to a particular choice of the channel basis in the leads. This means that we can perform a rotation $S 
\mapsto V^T S V$ with an arbitrary unitary matrix $V$ block-diagonal in the left-right channel space. For the minimal model with three channels, matrix $V$ 
contains five variables. Thus we conclude that an S-TS junction with $2 + 1$ channels is characterised by four parameters. We denote these parameters $T_1$, 
$T_A$, $\rho$, and $\chi$ and define them as
\begin{subequations}
\label{invariants}
\begin{gather}
 T_1 = t_{RL} t^\dagger_{RL}, \qquad
 T_A e^{i\chi} = - t_{RL} t^*_{LR}, \\
 \rho = \frac{1}{2} \mathop{\mathrm{tr}} \left[r_L (r_L^\dagger-r_L^*)\right].
\end{gather}
\end{subequations}
The parameter $T_1$ is the single non-zero transmission eigenvalue and coincides with normal-state conductance of the junction. The expression for $T_A 
e^{i\chi}$ can be understood as an amplitude of a fictitious Andreev reflection process where an electron from the right lead normally transmits into the left 
lead, gets Andreev reflected into a hole, which then transmits back into the right lead. This resembles the Andreev conductance amplitude of the junction when 
the right lead is in the normal state $\Delta_\text{top} = 0$. However $T_A$ does not contain multiple reflection processes while the actual zero-bias Andreev 
conductance takes the quantised value $G_A=2e^2/h$ due to multiple scattering.\cite{Wimmer, Dahlhaus, PikulinNazarov} The physical meaning of the parameter 
$\rho$ will be discussed below.

A straight-forward calculation of the determinant (\ref{detD}) for a $3\times3$ scattering matrix yields, up to 
$\varphi$-independent factors
\begin{multline}
D(\varphi,\omega)
 = (2 - T_1) \omega^2 + T_1\sqrt{(\omega^2 + \Delta_0^2)(\omega^2 + \Delta_\text{top}^2)} \\
   + T_A \Delta_0 \Delta_\text{top} \cos(\varphi - \chi) + \Delta_0^2 \rho.
 \label{Dminimal}
\end{multline}
Unitarity of $S$ ensures that $0 \leq T_A \leq T_1 \leq 1$ as well as $0 \leq \rho \leq 2 - T_1$.
With the determinant \eqref{Dminimal}, we obtain the Josephson current from Eq.\ \eqref{itodet},
\begin{equation}
I=\frac{e\Delta_0\Delta_{\mathrm{top}}}{ h}T_A\sin(\varphi-\chi)T\sum\limits_\omega \frac1{D(\varphi, \omega)}.
\label{I21}
\end{equation}
This general expression constitutes a central result of our paper.

Matsubara summation in Eq.\ \eqref{I21} can be transformed into an integral over real energies $E = i\omega$. This allows to identify contributions to the 
current from discrete and continuous parts of the spectrum of the junction. A discrete sub-gap Andreev bound state appears when the integrand has a pole at 
$E_0 < \min\{\Delta_0, \Delta_\text{top}\}$. The corresponding residue yields the supercurrent carried by this bound state. There is also the Majorana zero mode 
in the system, but it is insensitive to $\varphi$ and produces an $\omega$ factor in $D$, which was already dropped in Eq.\ \eqref{Dminimal}. Second, the 
integrand has a branch cut at $\min\{\Delta_0, \Delta_\text{top}\} < E < \max\{\Delta_0, \Delta_\text{top}\}$. Integration around this cut yields the current 
carried by the continuous spectrum of states extending into one of the superconductors. Continuum states above $\max\{\Delta_0, \Delta_\text{top}\}$ do not 
carry a supercurrent. 

\begin{figure}
\centering
\hspace*{-3pt}\includegraphics[width=0.485\textwidth]{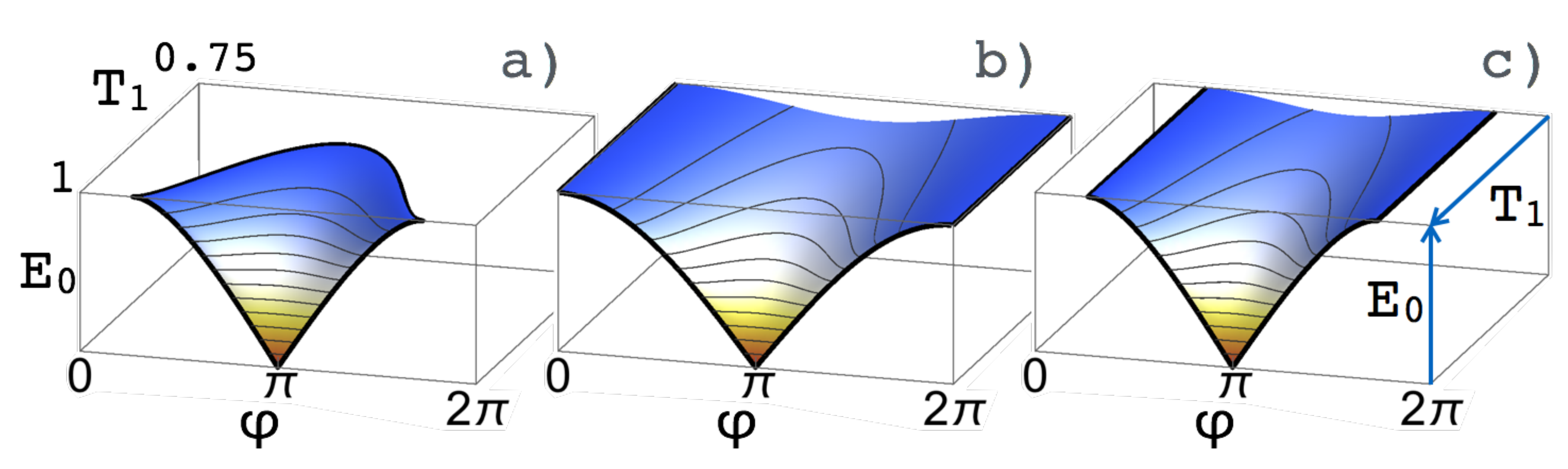}
\caption{(Color online) Bound state energy $E_0$ for the nanowire setup as a function of phase difference $\varphi$ and normal-state conductance $T_1$. The 
gaps are (a) $\Delta_0=1.5$, $\Delta_\text{top} = 1$, (b) $\Delta_0 = \Delta_\text{top} = 1$, and (c) $\Delta_0=1$, $\Delta_\text{top}=1.5$. Front cuts at 
perfect transmission $T_1=1$ are the same for (a) and (c). As $T_1$ decreases, the sub-gap state in case (a) quickly becomes shallow and disappears. In 
cases (b) and (c), the sub-gap state survives all the way to $T_1=0$ (note that the plots are shown only for $0.75 < T_1 < 1$). 
}
\label{fig2}
\end{figure}

Expression \eqref{I21} greatly simplifies when the two gaps are equal, $\Delta_0 = \Delta_\text{top}$. In this case, the current is given by Eq.\ 
\eqref{BeenakkerI} with a single current-carrying Andreev bound state at
\begin{align}
E_0 = \Delta \sqrt{\frac{\rho+T_1+T_A}{2}-T_A\sin^2\frac{(\varphi-\chi)}{2}}.
\label{boundstate}
\end{align}
For different gaps, $\Delta_0 \neq \Delta_{\mathrm{top}}$, the bound state energy $E_0$ is still straight-forward to calculate, but the resulting algebraic 
expression is much more cumbersome. It still has its minimum at $\varphi - \chi = \pi$, but the sub-gap level exists only in a range of phases around $\chi + 
\pi$ given by 
\begin{align}
\cos(\varphi-\chi) < \frac{(2-T_1)\min\{\Delta^2_0,\Delta^2_\text{top}\}}{T_A\Delta_0\Delta_{\mathrm{top}}}-\frac{\rho\Delta_0}{T_A\Delta_{\mathrm{top}}}.
\label{existence}
\end{align}
Equation \eqref{existence} inherits the asymmetry of Eq. \eqref{Dminimal} with respect to the gaps $\Delta_0$ and $\Delta_\text{top}$. In the case $\Delta_0 < 
\Delta_{\mathrm{top}}$, the r.h.s.\ of Eq.\ \eqref{existence} is proportional to $2 - T_1 - \rho > 0$, so that a sub-gap state exists at least in the interval 
$\pi/2 < \varphi < 3\pi/2$ where the cosine is negative. In the opposite case, $\Delta_0 > \Delta_\text{top}$ the bound state may be absent for all $\varphi$. 
The evolution of the Andreev state energy is illustrated in Fig.\ \ref{fig2}.

An important limit of Eq. \eqref{I21} is the tunneling limit $T_1 \ll 1$. In this case we get
\begin{align}
I=\frac{e\Delta_{\mathrm{top}}}{h\sqrt{2\rho}}T_A\sin(\varphi-\chi) \label{Itun}
\end{align}
provided $T_1 \ll \rho \Delta_0/\Delta_\text{top}$. A striking property of Eq.\ \eqref{Itun} is that the tunneling supercurrent does not depend on $\Delta_0$. 
This is in strong contrast to conventional junctions where the two leads and hence the two superconducting gaps enter on equal footing.\cite{Bagwell}

From Eqs. \eqref{I21} and \eqref{boundstate} we can extract the physical meaning of $\rho$. It becomes clear if we put both $T_1$ and $T_A$ to zero, which 
effectively disconnects the leads from each other. The bound state energy is $E_0 = \Delta_0\sqrt{\rho/2}$ in this case. Thus, $\sqrt\rho$ defines the energy 
of the sub-gap bound state at the end of the ordinary superconductor, when it is disconnected from the topological superconductor. If $\mathcal{T}$ symmetry is 
preserved in the terminated ordinary superconductor, then $r_L = -r_L^T$ leading to $\rho = 2$ so that there is no sub-gap level at the end of the lead.

The asymmetry of the results with respect to $\Delta_0 \leftrightarrow \Delta_\text{top}$ can be traced back to the broken time-reversal symmetry. In a 
conventional, $\mathcal{T}$-symmetric short junction the matrices $S$ and $S^T$ can be block-diagonalised simultaneously,\cite{Beenakker91} which decouples the 
contribution of different channels to the supercurrent. In the S-TS junction, $S$ and $S^T$ are not connected by $\mathcal{T}$-symmetry. Physically, this means 
that electrons and holes scatter differently at the junction -- if an electron is fully reflected, the corresponding hole might still be transmitted. This 
prevents us from eliminating a fully reflected electron channel from the problem. As a result, all three channels in the leads carry supercurrent in 
combination. The different number of channels associated with the two gaps explains the strong asymmetry of Eq.\ \eqref{I21} with respect to $\Delta_0$ and 
$\Delta_\text{top}$. 

\section{Nanowire setup}
\label{sec_nanowire}

Let us now apply the general results (\ref{Dminimal}) -- (\ref{I21}) to a particular model of a junction made of a nanowire with parabolic spectrum and strong 
spin-orbit coupling in the presence of an external magnetic field and proximity-induced superconductivity.\cite{Sarma, OregOppen} This model is widely used in 
literature to describe the existing experiments\cite{Mourik2012, Das2012, Deng2012} on the N-TS junctions.

The ordinary and topological parts of the junction differ by the material parameters or by doping (position of the chemical potential). We consider the model 
with a simple spin-degenerate left lead
\begin{align}
\xi_L = \frac{p^2}{2m} - \mu_L, \label{HL}
\end{align}
while the right lead is subject to strong spin-orbit coupling $u$ and parallel magnetic field $B$ (measured in energy units),
\begin{align}
\xi_R = \frac{p^2}{2m} - \mu_R + u p \sigma_z + B \sigma_x. \label{HR}
\end{align}
Here Pauli matrices $\sigma$ act in spin space. While the left lead represents a wire with two spin-degenerate conducting channels, the right lead hosts just a 
single channel if the chemical potential is inside the Zeeman gap, $|\mu_R| < |B|$. To describe the superconductivity induced in the wires, we use the 
Bogoliubov-de Gennes Hamiltonian (\ref{BdG}) with $\xi$ from the above two expressions. The right lead remains in the topological state\cite{Sarma, OregOppen} 
provided $|\Delta_{\mathrm{top}}| < \sqrt{B^2 - \mu_R^2}$. We will assume a stronger condition $\mu_R, \Delta_\text{top} \ll B \ll mu^2$. This will allow us 
to linearise the spectrum of the single channel close to the Fermi level and completely disregard the second gapped channel in the topological superconductor. 
It also establishes an approximate time-reversal symmetry with $\mathcal{T} = i \sigma_y \mathcal{C}$ in the right lead. 

Assuming a simple adiabatic point contact between the left and the right wire and matching the wave functions, we calculate the scattering matrix
\begin{align}
S=\begin{pmatrix}
\dfrac{4i\alpha}{(1+\alpha)^2} & \dfrac{1-\alpha}{1+\alpha} & \dfrac{2i(1-\alpha)\sqrt\alpha}{(1+\alpha)^2}\\
\dfrac{-1+\alpha}{1+\alpha} & 0 & \dfrac{2\sqrt\alpha}{1+\alpha}\\
\dfrac{2i(1-\alpha)\sqrt\alpha}{(1+\alpha)^2} & -\dfrac{2\sqrt\alpha}{1+\alpha} & \dfrac{i(1-\alpha)^2}{(1+\alpha)^2} 
\end{pmatrix}, \label{S3}
\end{align}
where the only parameter is the velocity ratio $\alpha=v_L/u$ with $v_L=\sqrt{2\mu_L/m}$. For this scattering matrix, the transport parameters are
\begin{gather}
T_1=\frac{8\alpha(1+\alpha^2)}{(1+\alpha)^4},\qquad T_A=\frac{16\alpha^2}{(1+\alpha)^4},\label{tta}\\
\rho=2\frac{(1-\alpha)^2}{(1+\alpha)^2},\qquad \chi=0.\label{rhochi}
\end{gather}
Substituting these values into Eq.\ \eqref{Itun}, we get the following result in the tunneling limit $\alpha \ll 1$:
\begin{align}
I=\frac{8\alpha^2 e\Delta_{\mathrm{top}}}{h}\sin\varphi. \label{Itun1}
\end{align}
At the same time, the normal-state resistance is $R_N = h/(e^2 T_1)= h/(8\alpha e^2)$. Thus we can express the critical current of the junction as
\begin{align}
I_c=\frac{h\Delta_{\mathrm{top}}}{8e^3R_N^2}.
\end{align}
We see that $I_c$ scales as inverse normal resistance squared, in strong contrast with the Ambegaokar-Baratoff relation \cite{Ambegaokar} which predicts 
linear scaling with inverse resistance. As mentioned earlier, $I_c$ depends on $\Delta_{\mathrm{top}}$ but not on $\Delta_0$ as long as $T_1 \ll 
\Delta_0/\Delta_{\mathrm{top}}$. In the case of equal gaps, the discrete level is $E_0=\Delta\sqrt{1-T_A\sin^2(\varphi/2)}$ which looks similar to Eq.\ 
\eqref{BeenakkerE} but is much more shallow in the tunnelling limit where $T_A\sim T_1^2$. 

Instead of using two different materials for the two leads, both superconductors can be made of the same nanowire with strong spin-orbit coupling. In this case, 
both leads are described by Eq.\ \eqref{HR} with different values of the chemical potential. In the ordinary (left) lead, we assume $\mu_L \gg B, mu^2$ which 
tunes the wire far into the topologically trivial phase. Since the two channels are only slightly split by spin-orbit interaction and the Zeeman effect, we 
can neglect these terms in the Hamiltonian and reduce the problem to the one studied above.

The peculiar supercurrent suppression leading to $I_c \sim \alpha^2$ is due to destructive interference of different processes contributing to the
supercurrent. The parameter $T_A$ describes the amplitude of an electron tunneling from the topological superconductor into the ordinary superconductor, then 
being Andreev reflected, followed by the tunneling of the hole back into the topological lead. This amplitude contains two terms, corresponding to the two 
intermediate states (channels or spin projections) that the particle can have in the ordinary superconductor. Each term scales as $\alpha \sim T_1$ in the 
tunneling limit, however, they interfere destructively with their sum scaling only as $\alpha^2 \sim T_1^2$.

\section{Junction with a random scattering matrix}
\label{sec_RMT}

The nanowire model considered in the previous Section exhibits a parametrically strong suppression of the Josephson current in the tunneling limit compared to 
what is expected from the Ambegaokar-Baratoff theory.\cite{Ambegaokar} In order to estimate how universal this suppression is, let us consider a generic 
junction with a random scattering matrix.

Since time-reversal symmetry is broken, we assume that $S$ is a random matrix from the circular unitary ensemble.\cite{Dyson} By calculating the joint 
distribution function $P(T_A, T_1, \rho, \chi)$ in this ensemble, we can find whether it is likely that a random junction has $T_A \ll T_1$ so that the 
supercurrent is parametrically small.

A unitary $3 \times 3$ matrix $S$ has nine independent parameters $x_{i}$ with $i = 1,\dots 9$. Let us assume that the first four of them are $T_1$, $T_A$, 
$\rho$, and $\chi$, which are important for transport. The five remaining parameters amount to rotations of the channel basis in the leads, $S = V^T S_0 V$. 
Here the matrix $S_0$ depends on the four transport parameters while $V$ is a block-diagonal (in left-right space) unitary matrix parameterised by $x_5,\dots 
x_9$. The metric tensor $M_{ij}$ of the circular unitary ensemble is defined by the uniform measure $\mathop{\mathrm{tr}} (dS\,dS^\dagger) = M_{ij}dx_idx_j$. 
The overall distribution function is $P(x_1,\dots x_9) = \sqrt{\det M}$. Integrating out the parameters $x_5,\dots x_9$, we obtain the distribution function for 
the relevant transport characteristics.

The distribution function takes a simple form in terms of the ratio $T_A/T_1$ instead of $T_A$ and an angular variable $\lambda \in (0, 2\pi)$ instead of 
$\rho$. This new variable is defined by
\begin{equation}
\rho = \left( 1 - \frac{T_A^2}{T_1^2} \right) \left( 1 - \frac{T_1}{2} + \sqrt{1 - T_1} \cos\lambda \right).
\end{equation}
An explicit calculation yields the following result:
\begin{align}
P\left( T_1,\frac{T_A}{T_1}, \chi, \lambda \right)
 = \mathrm{const}\; T_1\sqrt{\frac{2 - T_1}{1 - T_1}}\, \frac{T_A}{T_1}.
\end{align}
Here the constant factor provides proper normalisation. The distribution function does not depend on $\chi$ and $\lambda$ and factorises as a function of $T_1$ 
and $T_A/T_1$. Thus the four parameters are uncorrelated and the average $T_A$ for a given $T_1$ is
\begin{equation}
 \langle T_A \rangle = \frac{2 T_1}{3}.
\end{equation}
We conclude that typically $T_A \sim T_1$, so that $I_c \sim e\Delta_\text{top} / R_N$ in the tunneling limit and the Ambegaokar-Baratoff relation 
qualitatively holds.

Systems with $T_A \ll T_1$ have a very low probability within the circular unitary ensemble. The nanowire setup discussed in the previous Section represents 
one such rare realisation. This implies that a generic perturbation of the nanowire Hamiltonian may significantly increase the value of the critical current.

\section{Conclusion}
\label{sec_conclusion}

In conclusion, we have studied the Josephson effect in a junction between a topological and an ordinary (topologically trivial) superconductor. Within the 
quasi-one-dimensional model they are characterised by an odd and even number of channels, respectively. A topological superconductor inevitably breaks 
time-reversal symmetry at its end, which leads to the appearance of an unpaired Majorana mode at zero energy. We have derived general expressions 
(\ref{detD}), (\ref{ID}) for the supercurrent for a class of short topological junctions in terms of the normal state scattering matrix $S$. For the minimal 
model of the junction with two and one channels in the leads, the current (\ref{I21}) explicitly depends on the four invariants of the scattering matrix 
(\ref{invariants}). The result (\ref{I21}) is strongly asymmetric with respect to the two superconducting gaps $\Delta_0$ and $\Delta_\text{top}$.

We have also applied the general result to a model of a nanowire setup with parabolic spectrum and strong spin-orbit coupling in an external magnetic 
field and find a parametrically strong suppression of the critical current $I_c \propto \Delta_{\mathrm{top}}/R_N^2$ in the tunneling limit. However, this 
behaviour is not universal and a generic perturbation of the model parametrically enhances the critical current. This is established with the help of the 
random matrix theory applied to a generic junction with three channels.

We are grateful to Carlo Beenakker and Anton Akhmerov for valuable discussions. The work was supported by Russian Science Foundation (Grant No.\ 14-42-00044).


\begin{thebibliography}{99}

\bibitem{reviewQiZhang} 
X.-L.\ Qi, S.-C.\ Zhang, Rev.\ Mod.\ Phys.\ \textbf{83}, 1057 (2011).
\bibitem{reviewHasanKane} 
M.\ Z.\ Hasan, C.\ L.\ Kane, Rev.\ Mod.\ Phys.\ \textbf{82}, 3045 (2010).
\bibitem{Kitaev01} 
A.\ Yu.\ Kitaev, Physics-Uspekhi \textbf{44}, 131 (2001).
\bibitem{reviewTopComp} 
C.\ Nayak, S.\ H.\ Simon, A.\ Stern, M.\ Freedman, S.\ Das Sarma, Rev.\ Mod.\ Phys. \textbf{80}, 1083 (2008).

\bibitem{reviewAlicea} 
J.\ Alicea, Rep.\ Prog.\ Phys.\ \textbf{75}, 076501 (2012).
\bibitem{reviewBeenakker} 
C.\ W.\ J.\ Beenakker, Annu.\ Rev.\ Con.\ Mat.\ Phys. \textbf{4}, 113 (2013).

\bibitem{FuKaneZBP} L.\ Fu and C.\ L.\ Kane, Phys.\ Rev.\ Lett.\, \textbf{102}, 216403 (2009).
\bibitem{AkhmerovZBP} A.\ R.\ Akhmerov, J.\ Nilsson, and C.\ W.\ J.\ Beenakker, Phys.\ Rev.\ Lett., \textbf{102}, 216404 (2009).
\bibitem{LawLeeNg} K.\ T.\ Law, P.\ A.\ Lee, and T.\ K.\ Ng, Phys.\ Rev.\ Lett.,\ \textbf{103}, 237001 (2009).
\bibitem{IF2013}  
P.\ A.\ Ioselevich, M.\ V.\ Feigelman, New\ J.\ Phys.\ \textbf{15}, 055011 (2013).
\bibitem{IOF2012}  
P.\ A.\ Ioselevich, P.\ M.\ Ostrovsky, M.\ V.\ Feigelman, Phys.\ Rev.\ B \textbf{86}, 035441 (2012).

\bibitem{Mourik2012} V.\ Mourik, K.\ Zuo, S.\ M.\ Frolov, S.\ R.\ Plissard, E.\ P.\ A.\ M.\ Bakkers, and L.\ P.\ Kouwenhoven, Science \textbf{336}, 1003 (2012).
\bibitem{Das2012} A.\ Das, Y.\ Ronen, Y.\ Most, Y.\ Oreg, M.\ Heiblum, and H.\ Shtrikman, Nature Physics \textbf{8}, 887 (2012).
\bibitem{Deng2012} M.\ T.\ Deng, C.\ L.\ Yu, G.\ Y.\ Huang, M.\ Larsson, P.\ Caroff, H.\ Q.\ Xu, Nano Lett. \textbf{12}, 6414 (2012).
\bibitem{Sarma}  
R.\ M.\ Lutchyn, J.\ D.\ Sau, and S.\ Das Sarma, Phys.\ Rev.\ Lett.\ \textbf{105}, 77001 (2010).
\bibitem{OregOppen}  
Y.\ Oreg, G.\ Refael, F.\ von Oppen, Phys.\ Rev.\ Lett.\ \textbf{105}, 177002 (2010).

\bibitem{Yakovenko} H.-J.\ Kwon, K.\ Sengupta, V.\ M.\ Yakovenko, Low Temp.\ Phys.\ \textbf{30}, 613 (2004).
\bibitem{FuKane2009} L.\ Fu and C.\ L.\ Kane, Phys.\ Rev.\ B.  \textbf{79}, 161408 (2009).
\bibitem{IF2011} P.\ A.\ Ioselevich, M.\ V.\ Feigelman, Phys.\ Rev.\ Lett.\ \textbf{106}, 077003 (2011).
\bibitem{AliceaOppen} L.\ Jiang, D.\ Pekker, J.\ Alicea, G.\ Refael, Y.\ Oreg, and F.\ von
Oppen, Phys.\ Rev.\ Lett., \textbf{107}, 236401 (2011).
\bibitem{RanHosur}Y. Ran, P. Hosur, and A. Vishwanath, Phys. Rev. B 84, 184501 (2011).

\bibitem{Ambegaokar} V.\ Ambegaokar, A.\ Baratoff, Phys.\ Rev.\ Lett.\ \textbf{10}, 486 (1963).
\bibitem{Beenakker91} C.\ W.\ J.\ Beenakker and H.\ van Houten, Phys.\ Rev.\ Lett.\ \textbf{66}, 3056 (1991).
\bibitem{Bagwell} L.\ -F.\ Chang and P.\ F.\ Bagwell, Physical Review B, \textbf{49}, 22 (1994).
\bibitem{BrouwerBeenakker} P.\ W.\ Brouwer, C.\ W.\ J.\ Beenakker, Chaos, Solitons and Fractals \textbf{8}, 1249 (1997).
\bibitem{Wimmer} M.\ Wimmer, A.\ R.\ Akhmerov, J.\ P.\ Dahlhaus, and C.\ W.\ J.\ Beenakker, New J.\ Phys.\ \textbf{13} 053016 (2011).

\bibitem{Dahlhaus}  
A.\ R.\ Akhmerov, J.\ P.\ Dahlhaus, F.\ Hassler, M.\ Wimmer, and C.\ W.\ J.\ Beenakker, Phys.\ Rev.\ Lett.\ \textbf{106}, 057001 (2011).
\bibitem{PikulinNazarov}
D.\ I.\ Pikulin and Yuli V.\ Nazarov, JETP Letters, \textbf{94}, 9, 693-697 (2011).
\bibitem{Dyson} F.\ M.\ Dyson, J.\ Math.\ Phys.\ \textbf{3}, 1199 (1962).


\end{thebibliography}
\end{document}